\documentclass[aps,pre,reprint,superscriptaddress,nofootinbib,captions=justified]{revtex4-2}

\usepackage{amsmath,amssymb,graphicx,bm}
\usepackage{mathtools}
\usepackage{tikz}
\usepackage{txfonts}
\usepackage{hyperref}
\usepackage{enumitem}
\usepackage{url}
\usepackage{CJK}
\CJKnospace
\usetikzlibrary{arrows.meta,positioning}

\tikzset{
  node/.style={circle, draw, thick, minimum size=8mm, inner sep=0pt,
               font=\small},
  edge/.style={line width=1.1pt},
}

\begin{document}
\begin{CJK*}{UTF8}{mj}

\title{Friendship-paradox paradox: Do most people's friends really have more friends than they do?} 

\author{Sang Hoon Lee (이상훈)}
\email{lshlj82@gnu.ac.kr}
\affiliation{Department of Physics, Gyeongsang National University, Jinju, 52828, Korea}
\affiliation{Research Institute of Natural Science, Gyeongsang National University, Jinju 52828, Korea}
\affiliation{Future Convergence Technology Research Institute, Gyeongsang National University, Jinju 52849, Korea}

\date{\today}

\begin{abstract}
The classical friendship paradox asserts that, on average, an individual's
neighbors have a higher degree than the individual. This statement concerns
network-level means and does not describe how often a typical node is locally
dominated by its neighbors. Motivated by this distinction, we develop a
framework that separates mean-based friendship paradox inequalities from two majority-type quantities: a global fraction measuring how many nodes have a degree smaller than the mean degree of their neighbors, and a local fraction based on hub centrality that measures how many nodes are dominated in a median-based sense. We show that neither fraction is constrained by the classical friendship paradox and that they can behave independently of each other. A simple example and two
empirical networks illustrate how quadrant patterns in the joint distribution of
a node's degree and its neighbors' degree determine the signs and magnitudes of the two fractions, and how left- or right-skewed degree distributions of neighboring nodes can yield opposite conclusions for mean-based and median-based comparisons. The resulting framework offers a clearer distinction between population averages and local majority relations and provides a foundation for future analyses of local advantage, disadvantage, and perception asymmetry in complex networks.
\end{abstract}

\maketitle
\end{CJK*}

\section{Introduction}

The friendship paradox (FP), first articulated in Ref.~\cite{Feld1991}, is
usually summarized as the claim that ``your friends have more friends than you
do.'' This phrase has become a standard way to describe the bias that arises
when individuals compare their own connectivity with that of their neighbors in
a network. Behind this simple statement, however, lie several different ways to
formalize how one samples nodes and how one performs the comparison.
At the level of network-wide averages, the classical FP admits two natural
formulations. The first, sometimes called the alter-based version, focuses on
nodes reached by following randomly chosen edges and compares their average
degree with the population average. The second, the ego-based version, averages
for each node the degrees of its neighbors and then compares that average with
the mean degree over nodes. These two formulations coincide in degree-uncorrelated
networks but can differ when degree-degree correlations are present. Recent
work~\cite{Kumar2024,LeeFP2025_arXiv} has clarified how their difference is
controlled by correlations between a node's degree and the average degree of its
neighbors.

A separate but closely related family of questions asks not about averages over
the whole network, but about the fraction of nodes that are locally
disadvantaged relative to their neighbors. One global majority-type criterion
asks whether more than half of the nodes have a degree smaller than the mean
degree of their neighbors. A different, more local majority criterion uses the
hub centrality~\cite{Jeong2025PRE}, or the fraction-based peer pressure~\cite{ELee2019}, which
measures for each node the fraction of neighbors whose degree is smaller than
its own. In that framework, one can ask how often a node is outnumbered by
higher-degree neighbors in a median sense. These majority-type inequalities are
conceptually different from the original FP because they concern medians or
fractions of nodes rather than network-level means.

In practice, the colloquial phrasing of the friendship paradox is often taken to express a majority claim such as ``for most people, their friends have more friends than they do.'' This interpretation implicitly assumes that a typical individual is locally less connected than the people around them. The classical friendship paradox, however, makes no such statement: it concerns only network-wide averages and offers no guarantee about what happens for most nodes, or even for any specific node. A few earlier works---most notably those introducing and analyzing the \emph{strong friendship paradox}~\cite{Kooti2014,Wu2017,Lerman2016}---have pointed out that local majority-based comparisons can behave very differently from the traditional mean-based formulation. Yet the conceptual distinction between population-level averages and node-level majority relations remains underemphasized in much of the broader literature. As the examples in this paper illustrate, it is entirely possible for both classical FP inequalities to hold while a majority of nodes are in fact more connected than the mean of their neighbors, or conversely, for many nodes to appear locally disadvantaged even when global averages suggest otherwise. These discrepancies arise because local comparisons depend sensitively on the distributional shape of neighbor degrees around each node, not merely on global means. Recognizing this gap motivates the need for a clearer framework separating mean-based and majority-based notions of ``having more friends,'' which we develop in the remainder of the paper.

The purpose of this work is to separate these notions clearly and to provide a
minimal mathematical language for doing so. After introducing the basic
notation in the next section, we review the two classical FP inequalities and
then define two majority-type quantities: a global mean-based fraction that
encodes the question ``is it likely that your friends have more friends than
you?'' and a local median-based fraction, based on hub centrality, that encodes
the question ``is it likely that most of your neighbors have more friends than
you?'' We then analyze how these quantities behave in a toy example and in two
empirical networks, highlighting the structural reasons why they can
be different. The main goal is to offer a conceptually transparent framework that can support future quantitative work on local advantage, disadvantage, and perception asymmetry in complex networks. 

\section{Mathematical framework}

\subsection{Network notation}

Consider a simple undirected network with the adjacency matrix
$A=\{a_{ij}\}$, where $a_{ij}=1$ if nodes $i$ and $j$ are connected and
$a_{ij}=0$ otherwise. The degree of node $i$ is
\begin{equation}
k_i = \sum_j a_{ij} \,,
\end{equation}
and the neighbor set and neighbor-degree multiset are
\begin{equation}
\mathcal{N}(i)=\{j: a_{ij}=1\}  \,,
\qquad
\mathcal{K}_i=\{k_j : j\in\mathcal{N}(i)\} \,.
\end{equation}
The mean neighbor degree of node $i$ is
\begin{equation}
k_{nn}(i)
 = \frac{1}{k_i}\sum_{j\in\mathcal{N}(i)} k_j  \,,
\end{equation}
defined whenever $k_i>0$. Averages over nodes with respect to the empirical
degree distribution $p(k)$ are written as
\begin{equation}
\langle f(k)\rangle_{\mathrm{n}}
 = \sum_{k} f(k)\, p(k) \,,
\end{equation}
where the notational convention $\langle \cdots \rangle_\mathrm{n}$ follows Ref.~\cite{LeeFP2025_arXiv}.

\subsection{Classical alter-based and ego-based FP inequalities}

The alter-based FP considers the degree of a node reached by following a
uniformly random edge. The corresponding mean degree,
\begin{equation}
\langle k_{\mathrm{friend}} \rangle_{\mathrm{n}}
 = \frac{\langle k^2 \rangle_{\mathrm{n}}}{\langle k\rangle_{\mathrm{n}}} \ge \langle k\rangle_{\mathrm{n}} \,,
\label{eq:FPclassic_v1}
\end{equation}
The ego-based FP instead averages the per-node mean neighbor degree:
\begin{equation}
\langle k_{nn}\rangle_{\mathrm{n}}
 = \frac{1}{N}\sum_i k_{nn}(i) \ge \langle k\rangle_{\mathrm{n}} \,.
\label{eq:FPclassic_v2}
\end{equation}
Those are the original formulations presented in Ref.~\cite{Feld1991}. 
For completeness, we briefly outline how the relation between the two
averages arises.

The alter-based mean degree can be written as
\begin{equation}
\langle k_{\mathrm{friend}} \rangle_{\mathrm{n}}
 = \frac{\langle k^2\rangle_{\mathrm{n}}}{\langle k\rangle_{\mathrm{n}}} \,,
\end{equation}
which follows from the fact that sampling nodes via edges biases the
degree distribution by a factor proportional to $k$.
On the other hand, the ego-based mean is obtained by averaging the
mean neighbor degree
\begin{equation}
k_{nn}(i) = \frac{1}{k_i}\sum_{j} a_{ij} k_j \,,
\end{equation}
over all nodes,
\begin{equation}
\langle k_{nn} \rangle_{\mathrm{n}} = \frac{1}{N}\sum_i k_{nn}(i) \,.
\end{equation}
A key combinatorial identity connects these quantities.
Multiplying $k_{nn}(i)$ by $k_i$ and summing over all nodes yields
\begin{equation}
\sum_i k_i k_{nn}(i)
 = \sum_{i,j} a_{ij} k_j
 = \sum_i k_i^2 \,,
\end{equation}
where the second equality follows from the symmetry $a_{ij}=a_{ji}$.
Dividing by $N$ gives
\begin{equation}
\langle k k_{nn} \rangle_{\mathrm{n}} = \langle k^2\rangle_{\mathrm{n}} \,.
\end{equation}
Substituting this relation into the definition of
$\langle k_{\mathrm{friend}} \rangle_{\mathrm{n}}$ immediately yields
\begin{equation}
\langle k_{\mathrm{friend}}\rangle_{\mathrm{n}}
 = \frac{\langle k k_{nn} \rangle_{\mathrm{n}}}{\langle k\rangle_{\mathrm{n}}} \,.
\end{equation}
Subtracting $\langle k_{nn}\rangle_{\mathrm{n}}$ from both sides then gives
\begin{equation}
\langle k_{\mathrm{friend}}\rangle_{\mathrm{n}}
 - \langle k_{nn}\rangle_{\mathrm{n}}
 = \frac{1}{\langle k\rangle_{\mathrm{n}}}
   \operatorname{Cov}_{\mathrm{n}}(k,k_{nn}) \,,
\label{eq:FP_relation}
\end{equation}
where the node-based covariance $\operatorname{Cov_\mathrm{n}}(k,k_{nn}) = \langle k\,k_{nn}\rangle_\mathrm{n}
-\langle k\rangle_\mathrm{n} \langle k_{nn}\rangle_\mathrm{n}$ indicates the covariance of the degree of a node and the average degree of its neighbors.
This identity shows that the difference
between the alter-based and ego-based friendship paradoxes is governed
entirely by the covariance between a node's degree and the average degree
of its neighbors.

Equation~\eqref{eq:FP_relation} can be rewritten in an equivalent
moment-based form originally derived in Ref.~\cite{Kumar2024} and shown to be
analytically identical in Ref.~\cite{LeeFP2025_arXiv},
\begin{equation}
\langle k_{nn}\rangle_\mathrm{n} - \langle k_{\mathrm{friend}}\rangle_\mathrm{n} 
= \rho
\sqrt{
\left(\frac{\kappa_1 \kappa_3 - \kappa_2^2}{\kappa_1}\right)
\left(\kappa_{-1} - \kappa_1^{-1}\right)
} \,,
\label{eq:PNAS}
\end{equation}
where $\rho=\mathrm{Corr}_{\mathrm{e}}(D_D,D_O^{-1})$ is the Pearson correlation
coefficient, measured over oriented edges, between the degree $D_D$ of the
destination node and the inverse degree $D_O^{-1}$ of the origin node.
Here $\kappa_m = N^{-1}\sum_{i=1}^N k_i^m$ denotes the $m$th moment of the degree
distribution.
This representation makes explicit how the difference between the two
friendship paradox averages depends not only on degree heterogeneity (through
the degree moments $\kappa_m$) but also on edge-level correlations encoded in
$\rho$. While Eq.~\eqref{eq:FP_relation} emphasizes a node-level covariance
between $k_i$ and $k_{nn}(i)$, Eq.~\eqref{eq:PNAS} expresses the same quantity in
terms of edge-sampled statistics, thereby highlighting the role of
degree--degree correlations along edges. The equivalence between
Eqs.~\eqref{eq:FP_relation} and \eqref{eq:PNAS} thus clarifies that both
expressions capture the same underlying structural mechanism, viewed from
complementary node-based and edge-based perspectives.

Note that both inequalities in Eqs.~\eqref{eq:FPclassic_v1} and \eqref{eq:FPclassic_v2} are statements about \emph{averages} across the network and do not
refer to the fraction of nodes that are individually dominated by their
neighbors.

\subsection{Global mean-based majority fraction}

To formalize a global majority-type interpretation, define
\begin{equation}
\phi_{\mathrm{global}}
 = \frac{1}{N}
   \sum_{i=1}^{N} \mathbf{1}_{\{k_i < k_{nn}(i)\}}  \,,
\label{eq:phi_global}
\end{equation}
where $\mathbf{1}_{\{\cdot\}}$ denotes the indicator function, equal to $1$ if
the condition inside the braces is true and $0$ otherwise. The quantity
$\phi_{\mathrm{global}}$ represents the fraction of nodes whose degree is
smaller than the mean degree of their neighbors. The colloquial claim
``it is \emph{likely} that your friends have more friends than you'' corresponds to
the majority condition 
\begin{equation}
\phi_{\mathrm{global}}>\frac{1}{2} \,.
\label{eq:phi_global_condition}
\end{equation}

Crucially, the inequalities in Eqs.~\eqref{eq:FPclassic_v1} and
\eqref{eq:FPclassic_v2} do \emph{not} imply any constraint on
$\phi_{\mathrm{global}}$. Both $\phi_{\mathrm{global}}>1/2$ and
$\phi_{\mathrm{global}}<1/2$ are possible, even in networks where the classical
FP holds in both alter-based and ego-based forms. The AFB network example
discussed later provides an explicit case where
$\phi_{\mathrm{global}}<1/2$ despite the FP being satisfied.

\subsection{Local median-based majority fraction}

The median-based perspective introduced 
in the context of local hubs~\cite{Jeong2025PRE} is encoded by the hub
centrality
\begin{equation}
h_i
 = \frac{1}{k_i}
   \sum_{j\in\mathcal{N}(i)}
   \mathbf{1}_{\{k_j < k_i\}}  \,,
\label{eq:hub_centrality}
\end{equation}
which measures the fraction of neighbors whose degree is smaller than that of
node $i$. 
This quantity is closely related to the neighborhood median, but the relation
requires some care.  When neighbor degrees are treated as a multiset
$\mathcal{K}_i$, the condition 
\begin{equation}
h_i < \frac{1}{2} \,,
\label{eq:median_equiv}
\end{equation}
implies that fewer than half of the
neighbors have degree strictly smaller than $k_i$.  This ensures that at least
half of the neighbors have degree $\ge k_i$, which is consistent with the
inequality
\begin{equation}
k_i \le \mathcal{K}_i^{(1/2)} \,,
\end{equation}
where $\mathcal{K}_i^{(1/2)}$ denotes the median of the multiset
$\mathcal{K}_i$.  
In the absence of ties and for sufficiently large neighborhoods, this condition
effectively coincides with the strict comparison
\begin{equation}
k_i < \mathcal{K}_i^{(1/2)} \,,
\end{equation}
and the two formulations become asymptotically equivalent in the thermodynamic
limit.  
However, for finite graphs with ties or very small neighborhoods, the
equivalence does \emph{not} in general hold as a biconditional.  These boundary
effects do not arise from the median formalism itself but from the strict
inequality built into the definition of hub centrality, which treats
equal-degree neighbors as neither ``higher'' nor ``lower.''
With these qualifications, the interpretation remains clear: a node with
$h_i < 1/2$ is locally dominated by higher-degree neighbors in a majority
(median-based) sense, although the precise relationship with the median requires
attention to ties and small-neighborhood effects.

The corresponding population fraction is
\begin{equation}
\phi_{\mathrm{local}}
 = \frac{1}{N}
   \sum_{i=1}^{N}
   \mathbf{1}_{\{h_i < 1/2\}}  \,,
\label{eq:phi_local}
\end{equation}
which answers the question ``is it likely that most of your neighbors have more
friends than you?'', i.e., whether
\begin{equation}
\phi_{\mathrm{local}}>\frac{1}{2} \,,
\label{eq:phi_local_condition}
\end{equation}
holds or not. Neither $\phi_{\mathrm{local}}$ nor its comparison with
one-half is constrained by the classical FP. As the toy example and the two
empirical networks in later sections will show, $\phi_{\mathrm{global}}$ and $\phi_{\mathrm{local}}$
can take values on opposite sides of one-half and vary almost independently,
even though both are derived from comparisons between a node and its neighbors.

The local median-based formulation in Eqs.~\eqref{eq:hub_centrality} and \eqref{eq:median_equiv} is mathematically equivalent to the ``fraction-based peer pressure'' measure introduced in Ref.~\cite{ELee2019}, apart from the treatment of ties when $k_i = k_j$ (ties are excluded both in Eq.~\eqref{eq:hub_centrality} and in Ref.~\cite{ELee2019}). A conceptual difference arises in how the node-level quantities are aggregated. Ref.~\cite{ELee2019} defines a network-level peer pressure by averaging the local pressures over all nodes, whereas Eq.~\eqref{eq:phi_local} in the present work takes the fraction of nodes satisfying the median-based condition $h_i < 1/2$. This contrast leads to an interesting duality: Eq.~\eqref{eq:phi_global} represents a network-level fraction derived from local comparisons of mean values, while Ref.~\cite{ELee2019} constructs a network-level average derived from local comparisons of median values. Finally, Eq.~\eqref{eq:phi_local} is a combination of both: it represents a network-level fraction derived from local comparisons of median values.

In the sections that follow, these quantities are evaluated in a simple toy
network, in the Zachary's karate club (ZKC) network~\cite{Zachary1977}, and in the National Collegiate Athletic Association Division~I American football (AFB) network~\cite{GirvanNewman2002,football_data}, and their joint behavior is interpreted in terms of their local organization of joint distributions of $h_i$ and $k_{nn}$.

\section{Systematic Difference Between the Criteria}

The four quantities, $\langle k_{\mathrm{friend}}\rangle_{\mathrm{n}}$, 
$\langle k_{nn}\rangle_{\mathrm{n}}$, $\phi_{\mathrm{global}}$, and 
$\phi_{\mathrm{local}}$, encode two fundamentally different modes of comparison: mean-based relations at
the level of population averages, and median-based relations expressed as
majority conditions at the level of individual nodes. Although the classical FP
guarantees the two mean-based inequalities, no analogous constraint binds the
majority-type quantities. The mean and the median of the same neighbor-degree
multiset $\mathcal{K}_i$ can respond quite differently to local heterogeneity.
A single high-degree neighbor can substantially raise $k_{nn}(i)$ while leaving
$\mathcal{K}_i^{(1/2)}$ almost unchanged, and degree-degree correlations can
further amplify such discrepancies across the network.

Consequently, the two majority-type inequalities $\phi_{\mathrm{global}}$ and
$\phi_{\mathrm{local}}$ need not align with each other, nor with the classical
FP. Their values may lie above or below one-half depending on how degree values
and neighborhood structures are distributed. This naturally motivates the
question of how small or simple a network can be while already displaying such
divergence. The following toy example provides a simple construction in which
$\phi_{\mathrm{global}}<1/2$ and $\phi_{\mathrm{local}}<1/2$ hold
simultaneously, even though the classical FP inequalities remain satisfied.

\subsection{A toy example with $\phi_{\mathrm{global}} < 1/2$ and $\phi_{\mathrm{local}}<1/2$}

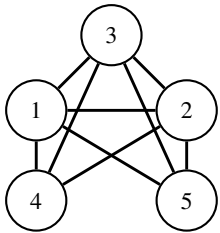
\begin{figure}[t]
  \centering
  \begin{tikzpicture}[scale=1.0]
    \node[node] (v1) at (-1,1) {1};
    \node[node] (v2) at (1,1) {2};
    \node[node] (v3) at (0,2) {3};
    \node[node] (v4) at (-1,-0.2) {4};
    \node[node] (v5) at (1,-0.2) {5};

    \draw[edge] (v1) -- (v2);
    \draw[edge] (v2) -- (v3);
    \draw[edge] (v3) -- (v1);

    \draw[edge] (v4) -- (v1);
    \draw[edge] (v4) -- (v2);
    \draw[edge] (v4) -- (v3);

    \draw[edge] (v5) -- (v1);
    \draw[edge] (v5) -- (v2);
    \draw[edge] (v5) -- (v3);
  \end{tikzpicture}
  \caption{%
  A five-node toy network with $\phi_{\mathrm{global}}=\phi_{\mathrm{local}}=2/5<1/2$.
  Nodes $1$, $2$, and $3$ form a triangle, and nodes $4$ and $5$ are connected
  to all of $1$, $2$, and $3$.  Degrees are $k_1=k_2=k_3=4$ and
  $k_4=k_5=3$.  The high-degree nodes $1$--$3$ are not locally dominated by
  their neighbors, whereas the low-degree nodes $4$ and $5$ are.}
  \label{fig:toy_local}
\end{figure}

Figure~\ref{fig:toy_local} presents a five-node toy network that explicitly
realizes the case where both $\phi_{\mathrm{global}}$ and $\phi_{\mathrm{local}}$ become less than $1/2$.
In this network, the degrees are
\begin{equation}
k_1=k_2=k_3=4 \,,
\qquad
k_4=k_5=3 \,.
\end{equation}
For nodes $1$, $2$, and $3$, the neighbor degrees are
$\{4,4,3,3\}$, so
\begin{equation}
k_{nn}(1)=k_{nn}(2)=k_{nn}(3)
 = \frac{4+4+3+3}{4}=3.5 < 4 \,,
\end{equation}
and the mean-based inequality $k_i<k_{nn}(i)$ fails at these nodes.
For nodes $4$ and $5$, the neighbor degrees are $\{4,4,4\}$, so
\begin{equation}
k_{nn}(4)=k_{nn}(5)
 = \frac{4+4+4}{3}=4 > 3 \,,
\end{equation}
and they satisfy $k_i<k_{nn}(i)$.  Hence
\begin{equation}
\phi_{\mathrm{global}}
 = \frac{1}{5}\bigl(
 0+0+0+1+1
\bigr)
 = \frac{2}{5}<\frac{1}{2} \,.
\end{equation}
For the median-based quantity, the hub centralities are
\begin{equation}
h_1=h_2=h_3
 = \frac{2}{4}=\frac{1}{2} \,,
\qquad
h_4=h_5
 = \frac{0}{3}=0 \,.
\end{equation}
Nodes $1$, $2$, and $3$ therefore have exactly half of their neighbors with
smaller degree and do not contribute to $\phi_{\mathrm{local}}$, whereas
nodes $4$ and $5$ satisfy $h_i<1/2$ and are counted as being locally
dominated in the median sense. Although the cases $h_i = 1/2$ lie precisely at the decision boundary, $k_i = 4$ is definitely a dominant one compared to the only other degree value $k_i = 3$ in this example, so we retain them in this form for illustrative clarity. Consequently,
\begin{equation}
\phi_{\mathrm{local}}
 = \frac{1}{5}\bigl(
 0+0+0+1+1
\bigr)
 = \frac{2}{5}<\frac{1}{2} \,.
\end{equation}
This example shows that it is easy to construct situations where a minority of nodes are locally dominated (in either the mean- or median-based sense). The toy network highlights that $\phi_{\mathrm{global}}$ and $\phi_{\mathrm{local}}$ can take values below one-half, despite the classical FP inequalities.

\section{Empirical Examples: Karate Club and American Football Networks}
\label{sec:empirical}

\begin{table}[t]
\caption{Numerical values of the four quantities
$\langle k_{\mathrm{friend}} \rangle_{\mathrm{n}}$, 
$\langle k_{nn} \rangle_{\mathrm{n}}$,
$\phi_{\mathrm{global}}$, and $\phi_{\mathrm{local}}$
for the toy network in Fig.~\ref{fig:toy_local}, the ZKC network~\cite{Zachary1977}, and the AFB network~\cite{GirvanNewman2002,football_data}.
The classical FP quantities are always larger than $\langle k\rangle_{\mathrm{n}}$, 
while the majority-type quantities 
$\phi_{\mathrm{global}}$ and $\phi_{\mathrm{local}}$ vary widely across networks, 
demonstrating their independence from mean-based FP conditions. The relation between the two classical FPs~\cite{LeeFP2025_arXiv} in Eq.~\eqref{eq:FP_relation} is also confirmed for all of the cases.}
\centering
\begin{ruledtabular}
\begin{tabular}{lccc}
 & Fig.~\ref{fig:toy_local} & ZKC & AFB \\
\hline
$N$ 
  & $5$ 
  & $34$
  & $115$ \\
$\langle k \rangle_{\mathrm{n}}$ 
  & $3.6$
  & $4.588235$ 
  & $10.660870$ \\
$\langle k_{\mathrm{friend}} \rangle_{\mathrm{n}}$ 
  & $11/3 \approx 3.67$ 
  & $7.769231$ 
  & $10.734095$ \\
$\langle k_{nn} \rangle_{\mathrm{n}}$ 
  & $37/10 = 3.7$
  & $9.610210$ 
  & $10.721979$ \\
$\operatorname{Cov_\mathrm{n}}(k,k_{nn})/ \langle k\rangle_\mathrm{n}$
 & $-1/30 \approx -0.03$
 & $-1.840979$
 & $0.012116$ \\
$\phi_{\mathrm{global}}$ 
  & $0.4$
  & $0.852941$ 
  & $0.434783$ \\
$\phi_{\mathrm{local}}$  
  & $0.4$
  & $0.705882$ 
  & $0.843478$ \\
\end{tabular}
\end{ruledtabular}
\label{tab:empirical_results}
\end{table}

\begin{figure*}[t]
  \centering
  \includegraphics[width=\textwidth]{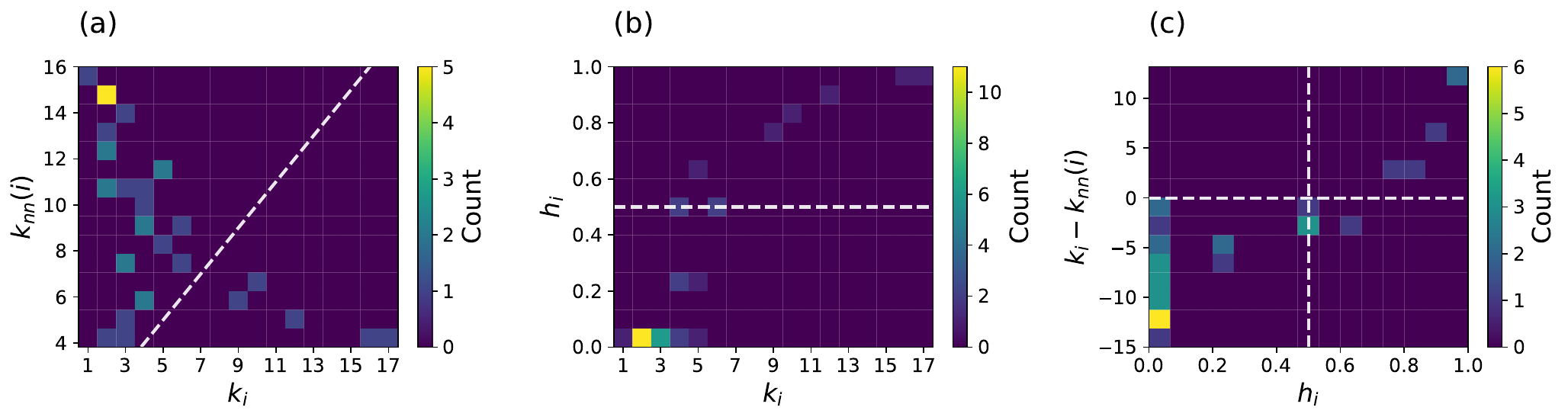}
  \caption{%
  Density plots for the ZKC network~\cite{Zachary1977},
  illustrating node-level relationships among degree $k_i$, mean neighbor
  degree $k_{nn}(i)$, hub centrality $h_i$, and the difference $k_i - k_{nn}(i)$ between a node's degree and the average value of its neighbors' degrees. The thresholds for the mean-based or median-based comparison, $k_{nn}(i)=k_i$ and $h_i = 1/2$, are plotted as the white dashed lines.
  Panel~(a) shows $k_i$ versus $k_{nn}(i)$, providing a node-level view of
  the ego-based FP; most points lie above the diagonal $k_{nn}(i)=k_i$, consistent with the
  large value $\phi_{\mathrm{global}} \approx 0.85$.  
  Panel~(b) shows $k_i$ versus $h_i$, emphasizing the median-based notion of
  local dominance.  
  Panel~(c) plots $h_i$ against $k_i - k_{nn}(i)$, revealing that nodes with
  negative mean-based contrast tend to have low hub centrality, indicating
  simultaneous mean-based and median-based disadvantage.  
  Together, the panels illustrate why both
  $\phi_{\mathrm{global}}$ and $\phi_{\mathrm{local}}$ are high in this
  network.}
  \label{fig:karate_density}
\end{figure*}
  
\begin{figure*}[t]
  \centering
  \includegraphics[width=\textwidth]{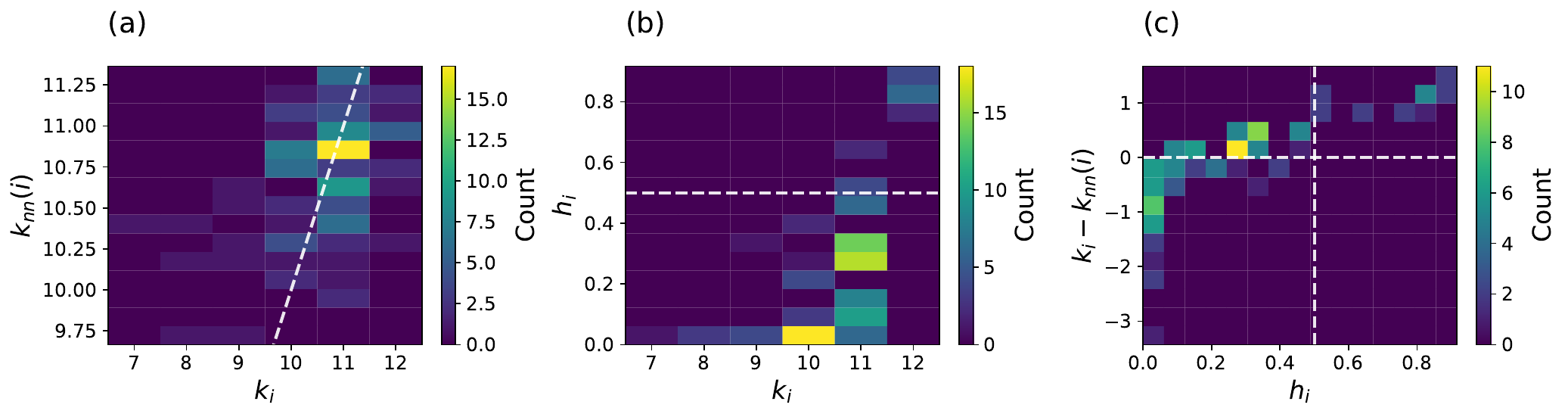}
  \caption{%
  Density plots for the AFB network
  \cite{GirvanNewman2002,football_data}, illustrating the same three
  relationships shown in Fig.~\ref{fig:karate_density}.  
  Panel~(a) displays $k_i$ versus $k_{nn}(i)$ and shows that a majority of points lie
  below the diagonal $k_{nn}(i)=k_i$, consistent with
  $\phi_{\mathrm{global}} \approx 0.43 < 1/2$, i.e., a majority of teams have degree larger than the mean degree of their neighbors, even though the classical FP in Eqs.~\eqref{eq:FPclassic_v1} and \eqref{eq:FPclassic_v2} holds.  
  Panel~(b) shows $k_i$ versus $h_i$, and panel~(c) shows $h_i$ versus
  $k_i - k_{nn}(i)$; both plots reveal that a large fraction of nodes satisfy
  $h_i < 1/2$, yielding $\phi_{\mathrm{local}} \approx 0.84$.  
  These contrasting patterns demonstrate that mean-based and median-based
  majority-type inequalities in Eqs.~\eqref{eq:phi_global_condition} and \eqref{eq:phi_local_condition} can also be quite different, in contrast to the toy network in Fig.~\ref{fig:toy_local} with $\phi_{\mathrm{global}} = \phi_{\mathrm{local}} < 1/2$.}
  \label{fig:football_density}
\end{figure*}

To illustrate the four measures discussed in this paper, consider two simple (and perhaps overused, but not in this context) empirical data: the ZKC network~\cite{Zachary1977} and the AFB network
compiled and published in Refs.~\cite{GirvanNewman2002,football_data}.
For each network, the two classical FP quantities
$\langle k_{\mathrm{friend}} \rangle_{\mathrm{n}}$ and
$\langle k_{nn} \rangle_{\mathrm{n}}$ are computed, along with the two
median-based fractions, $\phi_{\mathrm{global}}$ and $\phi_{\mathrm{local}}$, defined earlier in Eqs.~\eqref{eq:phi_global} and \eqref{eq:phi_local}.
Table~\ref{tab:empirical_results} summarizes the numerical values of the four quantities for all three networks.

The density plots in Figs.~\ref{fig:karate_density} and \ref{fig:football_density} provide a finer-grained view of how mean-based and median-based comparisons diverge at the node level. In the ZKC network (Fig.~\ref{fig:karate_density}),
panel~(a) shows that a large concentration of nodes lies above the diagonal
$k_{nn}(i)=k_i$, consistent with $\phi_{\mathrm{global}}\approx 0.85$: most individuals have a degree smaller than the mean degree of their neighbors.
Panel~(b) demonstrates that many of these same nodes have low hub centrality
$h_i$, and panel~(c) confirms that negative values of $k_i - k_{nn}(i)$ tend to
co-occur with low $h_i$.
The three panels together indicate that, for this network, mean-based and
median-based notions of being ``less connected than one's neighbors'' align
for most nodes, yielding simultaneously large values of
$\phi_{\mathrm{global}}$ and $\phi_{\mathrm{local}}$. 

In the AFB network (Fig.~\ref{fig:football_density}), however, panel~(a) shows a
substantial density of nodes below the diagonal, consistent with
$\phi_{\mathrm{global}}\approx 0.43<1/2$: the majority of teams have degree
larger than the mean degree of their neighbors.
Panels~(b) and (c), however, reveal a contrasting pattern: most nodes satisfy
$h_i<1/2$, producing $\phi_{\mathrm{local}}\approx 0.84$.
Thus, even though a majority of teams exceed the mean degree of their neighbors, they are still locally dominated in a median-based sense.
The juxtaposition of panels~(a)--(c) therefore illustrates how majority-type median-based inequalities can be different from mean-based comparisons, even when both are derived from the same neighborhood degree data.

Examining the distribution of points across the four quadrants of panel (c) in Figs.~\ref{fig:karate_density} and \ref{fig:football_density} offers an immediate visual explanation for the observed values of $\phi_{\mathrm{global}}$ and $\phi_{\mathrm{local}}$. In the ZKC network, most points fall in the first and third quadrants, where $k_i - k_{nn}(i)$ and $h_i - 1/2$ share the same sign. Nodes that are disadvantaged in the mean-based sense tend also to be disadvantaged in the median-based sense, which explains why $\phi_{\mathrm{global}}$ and $\phi_{\mathrm{local}}$ take similarly large values. The AFB network presents a contrasting pattern: a substantial mass in the second quadrant corresponds to nodes with $k_i > k_{nn}(i)$ but $h_i < 1/2$. These nodes have neighbors $\mathcal{N}_i$ whose distribution of degrees in $\mathcal{K}_i$ is sufficiently left-skewed that the mean is pulled below the median. This creates a mismatch between mean-based and median-based comparisons and results in $\phi_{\mathrm{global}} < 1/2$ but $\phi_{\mathrm{local}} \approx 0.84$. In the hypothetical contrasting case of a right-skewed distribution where the nodes in the fourth quadrant of panel (c) is abundant, which results in many nodes with $k_i < k_{nn}(i)$ but $h_i > 1/2$, it would be possible that $\phi_{\mathrm{global}} > 1/2$ and $\phi_{\mathrm{local}} < 1/2$. 

Taken together, these examples show that the two classical FP inequalities and the two majority-type inequalities capture qualitatively different structural features. The FP remains a statement about network-level \emph{averages}, whereas $\phi_{\mathrm{global}}$ and $\phi_{\mathrm{local}}$ describe \emph{majority-type} neighborhood relations. Their systematic divergence across networks underscores the need to treat these quantities as distinct diagnostics rather than interchangeable formulations of ``your friends have more friends than you.''

\section{Discussion and Outlook}

The analysis in this work shows that a full account of how a node compares with
its neighbors requires distinguishing four logically independent inequalities:
the classical alter-based FP, expressed by
$\langle k_{\mathrm{friend}}\rangle_{\mathrm{n}} \ge \langle k\rangle_{\mathrm{n}}$;
the classical ego-based FP, expressed by
$\langle k_{nn}\rangle_{\mathrm{n}} \ge \langle k\rangle_{\mathrm{n}}$;
the global majority-type condition based on mean neighbor degree,
$\phi_{\mathrm{global}}$ can be either $\ge 1/2$ or $< 1/2$; and the local majority-type condition based
on median neighbor degree, $\phi_{\mathrm{local}}$ can also be either $\ge 1/2$ or $< 1/2$. The first two
inequalities follow directly from the definitions of edge-based and node-based
averages and hold for any network. The latter two,
by contrast, are majority statements whose truth values are unconstrained by the
classical FP. They capture structurally distinct neighborhood relations and
should therefore be treated as different diagnostics rather than as reformulations
of the FP itself.

The explicit construction in Fig.~\ref{fig:toy_local} demonstrates that both $\phi_{\mathrm{global}}$ and
$\phi_{\mathrm{local}}$ can fall below one-half. The toy example highlights the
fact that mean-based and median-based comparisons capture different local
phenomena: a node may be dominated in the mean sense but not in the median
sense, or vice versa. The empirical results in
Table~\ref{tab:empirical_results} reinforce this point. In the ZKC
network, both $\phi_{\mathrm{global}}$ and $\phi_{\mathrm{local}}$ are high,
indicating a broad consistency between mean-based and median-based local
disadvantage. In contrast, the AFB network exhibits
$\phi_{\mathrm{global}}<1/2$ yet $\phi_{\mathrm{local}}\approx 0.84$, showing
that most teams are dominated by their neighbors in a median sense even though
a majority have degree exceeding the \emph{mean} of their neighbors' degrees.
The density plots in Figs.~\ref{fig:karate_density} and
\ref{fig:football_density} illustrate these divergences visually: the signed
contrast $k_i-k_{nn}(i)$ and the hub-centrality-based quantity $h_i$ need not be
concomitant.

Overall, the results make clear that the classical FP concerns 
\emph{population-level averages}, whereas the quantities 
$\phi_{\mathrm{global}}$ and $\phi_{\mathrm{local}}$ capture 
\emph{local majority relations}. The quadrant structure in panel (c) of 
Figs.~\ref{fig:karate_density} and \ref{fig:football_density} shows that these 
majority-type inequalities depend sensitively on the joint distribution of 
$\left( h_i, k_{nn}(i) \right)$ and on the shape of each node's neighbor-degree distribution. In particular, left- or right-skewed degree distributions of neighboring nodes---when the mean is significantly different from the median---produce systematic mismatches between mean-based and median-based comparisons. These effects cannot be inferred from the classical FP alone and underscore that the four inequalities considered here diagnose structurally distinct forms of local advantage or disadvantage.

Several directions for future research follow naturally from this perspective. 
One avenue is the analytical characterization of $\phi_{\mathrm{global}}$ and 
$\phi_{\mathrm{local}}$ in canonical random graph ensembles, including the 
configuration model and scale-free networks, with particular attention to the 
full distribution of $k_{nn}(i)$ rather than only its mean. A second direction 
is to investigate how assortativity~\cite{Newman2002}, community structures~\cite{Fortunato2022}, and core-periphery~\cite{Rombach2017} or rich-club organization~\cite{Colizza2006} shape both the four inequalities and the quadrant patterns observed in panel (c) of Figs.~\ref{fig:karate_density} and \ref{fig:football_density}. Extending the analysis to temporal or adaptive networks~\cite{Holme2012} is equally promising, since growth, rewiring, or behavioral feedback can shift local degree contrasts in ways that classical FP averages cannot capture. Finally, applications in systems where local perception asymmetry has 
behavioral or functional consequences---such as social influence, information 
diffusion, infrastructure robustness, or neuronal circuits---may benefit from 
explicitly distinguishing mean-based and median-based notions of local 
disadvantage.

The hope is that viewing the FP alongside its mean-based and median-based local
relatives will provide a flexible and interpretable language for describing how
nodes perceive their structural position in complex networks. These quantities
open the door to a more comprehensive theory of local comparison, extending the
conceptual scope of the friendship paradox beyond its traditional
mean-based foundations. In any case, the seemingly ``paradoxical'' situations in which a majority of nodes appear to ``violate'' the friendship paradox---what one might playfully call the friendship-paradox paradox (FPP)---are resolved within the framework developed here. The distinction between mean-based FP statements and majority-type, median-based inequalities clarifies why such phenomena arise and ensures that no genuine contradiction exists. 

\emph{Notes.}---After the first version of this paper was posted,
the author was kindly informed that the median-based majority condition
studied here has been previously introduced under the name
``strong friendship paradox'' in several earlier works~\cite{Kooti2014,Wu2017,Lerman2016}.
These studies also emphasized the role of distributional shape in
separating mean-based and median-based comparisons, and documented
empirical settings in which median-based dominance is widespread.
The author gratefully acknowledges these important precedents and
encourages readers to view the present contribution as complementary:
the goal here is to provide a unified mathematical formulation of the
mean-based and median-based quantities, clarify their logical
independence, and highlight conditions under which majority-type
inequalities diverge from the classical FP.

\section*{Epilogue: Learning from misunderstanding---classroom anecdote}

The author first encountered the distinction between mean-based and majority-type
comparisons in a particularly memorable moment while teaching a network science
course.
As part of a homework assignment, students were asked to examine the FP for the
AFB network data~\cite{GirvanNewman2002,football_data}.
One student proudly reported that the FP \emph{does not} hold in
this network, pointing to the fact that
$\phi_{\mathrm{global}}<1/2$: for a majority of teams, the mean degree of their neighbors is actually \emph{lower} than their own degree.
The conclusion might sound reasonable, but incorrect, as any reader who is reading this paper up to this point can easily tell.

What had happened was a misconception that the author suspects many readers may share
initially: the student had implicitly replaced the FP's comparison of
\emph{averages} with a comparison of \emph{majorities}.
The classical ego-based FP concerns Eq.~\eqref{eq:FPclassic_v2}, which compares
network-wide means and indeed holds for the football network.
The quantity $\phi_{\mathrm{global}}$, on the other hand, asks a completely
different question: whether the \emph{majority} of nodes individually satisfy
$k_i < k_{nn}(i)$.
There is no mathematical reason for this fraction to exceed one-half, and in the
AFB network it clearly does not.
This classroom episode served as a useful reminder that the colloquial phrasing
of the FP can tempt readers into thinking about ``most'' instead of ``average.''
The mismatch revealed by the AFB network provides a concrete motivation for
carefully separating mean-based FP statements from majority-type,
median-oriented interpretations and for studying the latter in their own right.

\acknowledgments
This research was supported by the National Research Foundation of Korea (NRF) under the grant RS-2021-NR061247. The author appreciates Krzysztof Sienicki for pointing out a subtle but important technical issue regarding the relationship between the condition $h_i<1/2$ and the neighborhood median, which helped improve the clarity and accuracy of the presentation. The author also thanks Kristina Lerman for kindly pointing out earlier works on the strong friendship paradox and for helpful correspondence that helped situate the present results within the broader literature.


\begin{thebibliography}{99}

\bibitem{Feld1991}
S.\,L. Feld,
Why your friends have more friends than you do,
Am. J. Sociol. \textbf{96}, 1464 (1991).

\bibitem{Kumar2024}
V. Kumar, D. Krackhardt, and S. Feld,
On the friendship paradox and inversity: A network property with applications to privacy-sensitive network interventions,
Proc. Natl. Acad. Sci. USA \textbf{121}, e2306412121 (2024).

\bibitem{LeeFP2025_arXiv}
S.\,H. Lee,
Two variants of the friendship paradox: The condition for inequality between them,
New Phys.: Sae Mulli \textbf{76}, 169 (2026).

\bibitem{Jeong2025PRE}
W. Jeong and U. Yu,
Critical phenomena and strategy ordering with a hub-centrality approach in the aspiration-based coordination game,
Chaos \textbf{31}, 093114 (2021).

\bibitem{ELee2019}
E. Lee, S. Lee, Y.-H. Eom, P. Holme, and H.-H. Jo,
Impact of perception models on friendship paradox and opinion formation,
Phys. Rev. E \textbf{99}, 052302 (2019).

\bibitem{Kooti2014}
F. Kooti, N.\,O. Hodas, and K. Lerman,
Network weirdness: Exploring the origins of network paradoxes,
in \textit{Proceedings of the International AAAI Conference on Weblogs and Social Media} \textbf{8}, 266 (2014).

\bibitem{Wu2017}
X.-Z. Wu, A.\,G. Percus, and K. Lerman,
Neighbor-neighbor correlations explain measurement bias in networks,
Sci. Rep. \textbf{7}, 5576 (2017).

\bibitem{Lerman2016}
K. Lerman, X. Yan, and X.-Z. Wu,
The ``majority illusion'' in social networks,
PLOS ONE \textbf{11}, e0147617 (2016).

\bibitem{Zachary1977}
W.\,W. Zachary,
An information flow model for conflict and fission in small groups,
J. Anthropol. Res. \textbf{33}, 452 (1977).

\bibitem{GirvanNewman2002}
M. Girvan and M.\,E.\,J. Newman,
Community structure in social and biological networks,
Proc. Natl. Acad. Sci. USA \textbf{99}, 7821 (2002).

\bibitem{football_data}
``Football'' network dataset, NetworkX documentation example,
\url{https://networkx.org/documentation/stable/auto_examples/graph/plot_football.html}
(accessed 16 November 2025).

\bibitem{Newman2002}
M.\,E.\,J. Newman, Assortative mixing in networks, Phys. Rev. Lett. \textbf{89}, 208701 (2002).

\bibitem{Fortunato2022}
S. Fortunato and M.\,E.\,J. Newman, 20 years of network community detection, 
Nat. Phys. \textbf{18}, 848 (2022).

\bibitem{Rombach2017}
P. Rombach, M.\,A. Porter, J.\,H. Fowler, and P.\,J. Mucha, Core-periphery structure in networks (revisited), SIAM Rev. \textbf{59}, 619 (2017).

\bibitem{Colizza2006}
V. Colizza, A. Flammini, M.\,A. Serrano, and A. Vespignani,
Detecting rich-club ordering in complex networks,
Nat. Phys. \textbf{2}, 110 (2006).

\bibitem{Holme2012}
P. Holme and J. Saram{\"a}ki,
Temporal networks,
Phys. Rep. \textbf{519}, 97 (2012).

\end{thebibliography}
\end{document}